\global\def\draftcontrol{0}

   \def\versionno{ string solitons -- draft   }

\catcode`\@=11

\expandafter\ifx\csname draftcontrol\endcsname\relax\global\def\draftcontrol{0}
\fi

{\count255=\time\divide\count255 by 60
\xdef\hourmin{\number\count255}
\multiply\count255 by-60\advance\count255 by\time
\xdef\hourmin{\hourmin:\ifnum\count255<10 0\fi\the\count255}}
\def\draftdate{\number\month/\number\day/\number\year\ \ \ \hourmin }

\newcommand\makepapertitle{\par
  \begingroup
    \renewcommand\thefootnote{\@fnsymbol\c@footnote}%
    \def\@makefnmark{\rlap{\@textsuperscript{\normalfont\@thefnmark}}}%
    \long\def\@makefntext##1{\parindent 1em\noindent
            \hb@xt@1.8em{%
                \hss\@textsuperscript{\normalfont\@thefnmark}}##1}%
     \newpage
     \global\@topnum\z@   
     \@makepapertitle
     \thispagestyle{empty}\@thanks
  \endgroup
  \setcounter{footnote}{0}%
  \global\let\thanks\relax
  \global\let\makepapertitle\relax
  \global\let\@makepapertitle\relax
  \global\let\@thanks\@empty
  \global\let\@author\@empty
  \global\let\@date\@empty
  \global\let\@title\@empty
  \global\let\title\relax
  \global\let\author\relax
  \global\let\date\relax
  \global\let\and\relax
  \def\version{\let\version\@version\@gobble}
}
\def\@makepapertitle{%
  \newpage
   \ifnum\draftcontrol=1 {}
   \version\versionno
   \vskip 3em%
   \else
   \hfill\hbox to 3cm {\parbox{4cm}{\@pubnum}\hss}%
   \vskip 3em%
   \fi
   \begin{center}%
   \let \footnote \thanks
     {\LARGE {\@title}}%
     \vskip 1.5em%
     {\normalsize
       \lineskip .5em%
       \begin{tabular}[t]{c}%
         \@author
       \end{tabular}\par}%
     \vskip 1.5em%
     {\@bstract}%
     \end{center}%
     \vskip 1.5em 
     \@date%
   \par
}

\gdef\@pubnum{}
\def\pubnum#1{%
  \gdef\@pubnum{#1}}

\gdef\@bstract{}
\def\Abstract#1{%
  \gdef\@bstract{%
   \parbox{\textwidth-0pc}{%
   \centerline{\bf Abstract}\penalty1000
   \noindent
   \renewcommand\baselinestretch{1.0}
   {#1}}}
}

\def\ps@paper{\let\@mkboth\@gobbletwo%
     \ifnum\draftcontrol=1
	\def\@oddfoot{\hbox to \textwidth{\tiny \versionno \hfil\tiny\draftdate}%
	\hskip -\textwidth \hbox to \textwidth{\hfil\rm\thepage\hfil}}%
     \else\def\@oddfoot{\hbox to \textwidth{\hfil\rm\thepage\hfil}}
     \fi
     \let\@evenfoot\@oddfoot
}

\def\body{\clearpage
          \pagestyle{paper}
	}

\def\@version#1{\ifnum\draftcontrol=1
\typeout{}\typeout{#1}\typeout{}
\vskip3mm\centerline{\hbox{\fbox{\normalsize{\tt DRAFT -- #1 -- }
                   {\draftdate}}}}\vskip3mm
\fi}
\let\version\@version
\long\def\eqlabel#1{\ifnum\draftcontrol=1
                    \tag@false  
                    \tag*{(\theequation) \hbox to -0.2cm{\hspace{0cm}\small{#1}\hss}}
                    \refstepcounter{equation} 
                    \edef\@currentlabel{\theequation}
                    \ltx@label{#1}          
                    \else
                    \label{#1}
                    \fi
                    }
\let\st@bibitem\@bibitem
\let\st@lbibitem\@lbibitem
\ifnum\draftcontrol=1
  \def\@bibitem#1{%
    \st@bibitem{#1}\a@@label{#1}\ignorespaces}
  \def\@lbibitem[#1]#2{%
    \st@lbibitem[#1]{#2}\a@@label{#2}\ignorespaces}
  \def\a@@label#1{%
    \gdef\a@lab{\smash{\normalfont\small#1}}
    \ifvmode
      \if@inlabel
        \global\setbox\@labels\hbox{%
          \llap{\a@lab\let\a@lab\relax
                \kern\@totalleftmargin\kern\marginparsep}%
          \box\@labels}%
      \fi
    \fi}
\fi

\documentclass[12pt,letterpaper]{article}

\usepackage{amsmath,amssymb,array,calc,rotating,epsfig,psfrag}
\usepackage[nosort]{cite}

\ifnum\draftcontrol=1
\fi

\renewcommand\baselinestretch{1.25}
\renewcommand\section{\@startsection {section}{1}{\z@}%
                                   {-3.5ex \@plus -1ex \@minus -.2ex}%
                                   {2.3ex \@plus.2ex}%
                                   {\normalfont\large\bfseries}}
\renewcommand\subsection{\@startsection{subsection}{2}{\z@}%
                                   {-3.25ex\@plus -1ex \@minus -.2ex}%
                                   {1.5ex \@plus .2ex}%
                                   {\normalfont\normalsize\bfseries}}
\renewcommand\subsubsection{\@startsection{subsubsection}{3}{\z@}%
                                   {-3.25ex\@plus -1ex \@minus -.2ex}%
                                   {1.5ex \@plus .2ex}%
                                   {\normalfont\normalsize\it}}
\renewcommand\paragraph{\@startsection{paragraph}{4}{\z@}%
                                   {-3.25ex\@plus -1ex \@minus -.2ex}%
                                   {1.5ex \@plus .2ex}%
                                   {\normalfont\normalsize\bf}}

\def\ie{{\it i.e.}}

\def\revise#1       {\raisebox{-0em}{\rule{3pt}{1em}}%
                     \marginpar{\raisebox{.5em}{\vrule width3pt\
                     \vrule width0pt height 0pt depth0.5em
                     \hbox to 0cm{\hspace{0cm}{%
                     \parbox[t]{4em}{\raggedright\footnotesize{#1}}}\hss}}}}

\newcommand\nxt[1]  {\\\fnxt#1}

\def\calf         {{\cal F}}
\def\calg         {{\cal G}}

\def\caln         {{\cal N}}
\def\calo         {{\cal O}}

\def\zet          {{\mathbb Z}}

\def\del          {\partial}

\def\tr           {\mathop{\rm Tr}}

\def\sqr#1#2{{\vcenter{\vbox{\hrule height.#2pt  
 \hbox{\vrule width.#2pt height#1pt \kern#1pt
 \vrule width.#2pt}\hrule height.#2pt}}}}

\def\a{\alpha}
\def\e{\eta}
\def\b{\beta}
\def\r{\rho}
\def\s{\sigma}
\def\f{\phi}
\def\w{\omega}
\def\t{\tau  }
\def\dd{\delta}
\def\ep{\epsilon}
\def\la{\lambda}
\def\ga{\gamma}

\catcode`\@=12

\begin{document}


\title{Gauge/string correspondence in curved space}

\pubnum{%
MCTP-02-59 \\
hep-th/0211141}
\date{November 2002}

\author{Alex Buchel\\[0.4cm]
\it Michigan Center for Theoretical Physics \\
\it Randall Laboratory of Physics, The University of Michigan \\
\it Ann Arbor, MI 48109-1120 \\[0.2cm]
}

\Abstract{
We discuss Gubser-Klebanov-Polyakov proposal for the gauge/string theory
correspondence for gauge theories in curved space. Specifically,
we consider Klebanov-Tseytlin cascading gauge theory compactified on
$S^3$. We explain regime when this gauge theory is a small deformation 
of the superconformal $\caln=1$ gauge theory on the world volume 
of regular D3-branes at the tip of the conifold. We study closed 
string states on the leading Regge trajectory in this background, and 
attempt to identify the dual gauge theory twist two operators. 
}


\makepapertitle

\body

\version\versionno

\section{Introduction}
Probably the most intriguing aspect of the gauge theory/string theory 
duality \cite{m9711,bmn0202,gkp0204} 
(see \cite{a9905} for a review) is the fact that it provides 
a dynamical principle for the nonperturbative definition of string theory
in the asymptotically Anti de Sitter spacetime where there is no notion of an 
$S$-matrix\footnote{This is emphasized 
in particular in \cite{w0106}.}. It thus appear promising 
that this dual ``definition'' of string theory (in terms of certain 
non-gravitational gauge theory) will be useful for formulating string 
theory in cosmologically relevant backgrounds\footnote{Problems of 
defining $S$-matrix in backgrounds with observer dependent 
horizons are discussed in \cite{w0106,h0104,f0104}.}.    
Following on ideas of  constructing the supergravity dual to gauge 
theories in curved space-time \cite{bt}, it was proposed 
in \cite{b,blw} that certain gravitating de Sitter backgrounds of string 
theory are dual to gauge theories formulated in classical,
non-gravitational de Sitter space-time. 

Most work on the Maldacena proposal\footnote
{See citation to\cite{m9711}.}, including the specific computations
in \cite{bt,b,blw}, dealt with the supergravity limit 
(corresponding to the one-loop approximation of the sigma model) of the 
gauge/string theory correspondence. It has been recently explained in 
\cite{bmn0202,gkp0204} how  to identify string states in the 
dual gauge theory. Specifically, it has been argued in \cite{gkp0204} (GKP)
that  certain $\caln=4$ supersymmetric $SU(N)$
gauge theory states with large quantum numbers are described by 
solitons of the nonlinear sigma model in $AdS_5\times S^5$. 
Motivated mainly by the potential application of the 
deformations of gauge/string theory correspondence, where the gauge theory 
is formulated in curved space-time toward  observable cosmology,
in this paper we attempt to extend analysis of \cite{gkp0204} 
for (static) deformations of \cite{bt}.     
More precisely, we study\footnote{
GKP proposal for nonconformal gauge/string 
theory correspondence has been discussed in \cite{other}. 
} closed string states on the leading Regge 
trajectory in the supergravity background \cite{bt},
dual to the Klebanov-Tseytlin (KT) cascading gauge theory 
\cite{kt0002}, formulated in $R\times S^3$. 
Somewhat surprisingly, 	in the regime 
where we can trust both the gauge theory and the sigma model 
analysis, we find that natural 
candidate dual twist two operators of the gauge theory 
have subleading correction to the 
anomalous dimensions different from the corresponding 
correction to the energy of the highly excited closed
string states.

The paper is organized as follows. In the next section we 
briefly review the correspondence of \cite{gkp0204} and 
explain, from the gauge theory perspective, why the 
KT gauge theory in $R\times S^3$ is a natural (and 
computationally controllable) deformation of the 
GKP computation where one breaks the conformal invariance.
We perform the computation of the anomalous dimension of 
certain twist two operators  in this gauge theory near the 
infrared conformal fixed point. 
These are the operators analogous to the ones
identified in \cite{gkp0204} as the gauge theory dual to 
highly excited (``long'') strings on the leading Regge
trajectory in $AdS_5$. 
In section 3 we 
review and clarify the dual supergravity background 
originally constructed in \cite{bt}. We then study 
closed strings spinning in this background, and find a  
discrepancy  with the gauge theory computation of section 2. 
The disagreement (at the technical level) can be traced back  
to the fact  (mentioned in \cite{bt}) that the
leading correction to the global  $AdS_5\times T^{1,1}$
background due to the 3-form flux does not have the 
precise asymptotics of the extremal Klebanov-Tseytlin 
background.  We do not have a physical understanding 
of the discrepancy at this stage. 
We comment on the difficulties of 
establishing gauge/string correspondence for 
the  KT gauge theory in the far ultraviolet,
where it is also  believed to be almost conformal.
Since the bulk of the paper is rather technical, 
in section 4 we summarize the main logical steps,
avoiding formulas, leading to the puzzle.

\section{The gauge theory story}

The original Maldacena correspondence \cite{m9711} 
relates $\caln=4$ $SU(N)$ superconformal Yang-Mills  theory in 
four dimensional Minkowski space-time and type 
IIB string theory in $AdS_5\times S^5$. 
The gauge theory description is valid for small 
't Hooft coupling $\la\equiv g_{YM}^2 N\ll 1$,
while the dual supergravity has small curvatures
in the opposite regime $\la\gg 1$. The main challenge in 
extending the correspondence beyond the supergravity approximation 
comes from the fact that  anomalous dimensions of the 
gauge theory operators dual to excited string states
(rather than to the supergravity modes) are generically expected 
to grow as $\la^{1/4}$ \cite{gkp,w98}, and thus appear to be beyond the grasp 
of the perturbative gauge theory analysis in the regime of the 
validity of the supergravity approximation. 

\subsection{GKP proposal}

In \cite{gkp0204}, Gubser, Klebanov and Polyakov 
considered twist two operators in $\caln=4$ SYM. 
In the free field theory these are the operators with the 
lowest conformal dimension for a given spin $n$, for example,
\begin{equation}
\calo_{(\mu_1\cdots\mu_n)}=\tr \Phi^I\nabla_{(\mu_1}\cdots
\nabla_{\mu_n)|}\Phi^I\, ,  
\eqlabel{o} 
\end{equation}
where $\Phi^I$ are the $\caln=4$ scalars, and the symmetrization 
$(\cdots )|$ denotes also a removal of traces.  The gauge covariant 
derivative is $\nabla_{\mu}\equiv \del_{\mu}+i g_{YM} A_{\mu}$.
Classically, the operator $\calo_{(\mu_1\cdots\mu_n)}$ 
has dimension $\triangle_{\calo_n}=n+2$, hence twist 2. In is expected 
\cite{kor}, that the leading term in the anomalous dimensions 
of operators such as \eqref{o}  grows as $\ln n$ (exactly as a one-loop
perturbative correction!) to all orders in perturbation 
theory and also non-perturbatively. 
Thus, one expects that in the full interacting gauge theory 
\begin{equation}
\triangle_{\calo_n}-(n+2)=f(\la)\ln n+o(\ln n)\,,
\eqlabel{qo}
\end{equation}
where $f(\la)$ is a certain function of the 
't Hooft coupling, which  has perturbative SYM expansion
\begin{equation}
f(\la)=a_1\la+a_2\la^2+\cdots,\qquad (\la\ll 1)\,.
\eqlabel{deff}
\end{equation}
Notice that, provided \eqref{qo} is correct,
the anomalous dimension of operators $\calo_n$ is small compare to the 
classical one, in the limit $n\to \infty$
\begin{equation}
\frac{\triangle_{\calo_n}-(n+2)}{n}\to 0\,.
\eqlabel{ratio}  
\end{equation}
GKP proposed \cite{gkp0204} that such operators in the dual 
supergravity picture are described by folded macroscopic strings rotating 
in $AdS_5$. Remarkably, closed strings 
on the leading Regge trajectory with large spin $S\gg 1$ in the $AdS_5$  
have energy 
\begin{equation}
E=S+\frac{\sqrt{\la}}{\pi}\ln S+o(\ln S)\,, 
\eqlabel{elong}
\end{equation}
which agrees (up to the functional dependence on  $\la$)
with the gauge theory result \eqref{qo}, 
once we employ the standard gauge/gravity 
dictionary 
\begin{equation}
E\leftrightarrow \triangle,\qquad S\leftrightarrow n\,.
\eqlabel{dic}
\end{equation}
Notice that GKP proposal  predicts that
the strong coupling expansion of $f(\la)$ in \eqref{qo}
is\footnote{The 1-loop string correction $\tilde{a}_1$ 
has been computed in \cite{ft}.} 
\begin{equation}
f(\la)=\frac{\sqrt{\la}}{\pi}+\tilde{a}_1+\frac{\tilde{a}_2}{\sqrt{\la}}+
O(\frac{1}{\la})\,.
\eqlabel{flarge}
\end{equation}

\subsection{KT gauge theory on $S^3$}

We would like to extend GKP analysis to KT cascading gauge theory 
compactified on $S^3$. We explain the regime in which this theory 
is a ``small deformation''  of the Klebanov-Witten  (KW) $\caln=1$ 
superconformal gauge theory \cite{kw}, and estimate leading correction 
to the anomalous dimensions of operators analogous to \eqref{o}. 

Consider $N$ D3-branes at the tip of the singular conifold 
in type IIB string theory \cite{kw}. In the limit of small 
't Hooft coupling\footnote{$g_s$ is the string coupling.} 
$g_s N\ll 1$, the gauge theory on the world volume 
of the D-branes is weakly coupled. We find $\caln=1$ 
supersymmetric $SU(N)\times SU(N)$
gauge theory with two chiral superfields $A_1, A_2$ in the 
$(N,\overline{N})$ representation, and two fields $B_1, B_2$ in 
the $(\overline{N},N)$ representation. Additionally, there is 
a superpotential 
\begin{equation}
W\sim {\rm tr}\ (A_i B_j A_k B_{\ell}) \epsilon^{ik}\epsilon^{j\ell}\,.
\eqlabel{sup}
\end{equation}
The theory has $SU(2)\times SU(2)\times U(1)$ global symmetry,
with the first (second) $SU(2)$ factor rotating the flavor index 
of the $A_i$ $(B_i)$, while the ``baryon'' $U(1)$ acts as 
$A_i\rightarrow A_i e^{i\alpha}$, $B_i\rightarrow B_i e^{-i\a}$.
There is also anomaly free $U(1)_R$ symmetry 
under which $A_i, B_j$ superfields have $R$-charge\footnote{We assign 
$R$-charge to gauginos to be $1$.} $\frac 12$. 
As argued in \cite{ls}, this theory flows in the 
IR to a superconformal fixed point with the same $U(1)_R$ 
symmetry, and hence exactly marginal superpotential \eqref{sup}.
At the IR fixed point, the theory has two  exactly 
marginal deformations, parameterized by the gauge couplings 
$g_1,g_2$ of the $SU(N)\times SU(N)$.

The dual supergravity description of the IR fixed point 
of the above gauge theory is represented by the backreaction of the 
D3-branes on the conifold geometry in the limit $g_s N\gg 1$.
In this case one finds \cite{kw} $AdS_5\times T^{1,1}$ 
($T^{1,1}=(SU(2)\times SU(2))/U(1)$) background with $N$
units of the five form flux through the $T^{1,1}$. 
The $AdS_5$ factor reflects the conformal invariance 
of the dual gauge theory.  
The string coupling 
is constant, and is related to the sum of the two gauge couplings
\begin{equation}
\frac{1}{g_s}=\frac{4\pi}{g_1^2}+\frac{4\pi}{g_2^2}\,.
\eqlabel{sum}
\end{equation}  
The other modulus, the difference of the gauge couplings, 
is related to a (constant) NSNS 2-form flux $B_2$ 
through the $S^2$ of the cone base\footnote{Topologically 
$T^{1,1}$ is $S^2\times S^3$.}. 
Finally, the  symmetries of the $T^{1,1}$ coset are 
realized as global symmetries of the gauge theory.   

One can literally repeat the GKP analysis in this case. 
The energy-spin relation for the macroscopic rotating strings 
with $S\gg 1$ is still given by \eqref{elong}, while the dual 
gauge theory twist two operators are 
\begin{equation}
\calo^{ij}_{(\mu_1\cdots\mu_n)}={\rm tr} A_i\nabla_{(\mu_1}\cdots
\nabla_{\mu_n)|} B_j\,, 
\eqlabel{okw}
\end{equation}
with $\nabla_\mu\equiv\del_\mu+i g_1 A_\mu^{(1)}+i g_2 A_\mu^{(2)}$.
It is easy to see that the one-loop perturbative contribution 
to the anomalous dimension scales as
\begin{equation}
\triangle_{\calo_n^{ij}}-(n+2)\sim N (g_1^2+g_2^2) \ln n\,.
\eqlabel{okwa}
\end{equation}
Interestingly, while the perturbative gauge theory 
computation \eqref{okwa} indicates the dependence on 
both parameters along exactly marginal deformations
$g_1,g_2$, the corresponding classical Regge trajectory
depends only on the 't Hooft parameter 
\begin{equation}
\la\equiv g_s N=\frac{g_1^2 g_2^2 N}{4\pi(g_1^2+g_2^2)}\,.
\eqlabel{lamdef}
\end{equation}  
The obvious reason is that the classical string rotating in 
$AdS_5$ does not ``know'' about the NSNS 2-form flux through
the two-cycle of the  $T^{1,1}$, parameterizing  the other marginal direction 
\begin{equation}
\int_{S^2} B_2 \sim \frac{1}{g_1^2}-\frac{1}{g_2^2}\,.
\eqlabel{diffcoupl}
\end{equation}
We expect however, that the dependence on \eqref{diffcoupl}
would arise in 1-loop sigma model correction to the 
Regge trajectory. It would be very interesting 
to verify this explicitly by extending the computation of 
\cite{ft}. The changes in the $\la$ scaling from the 
weak to strong coupling regime is familiar from 
other gauge/gravity computations (like the coefficient 
of Wilson loops), or the GKP analysis.   
The coupling dependence we are finding here is much more involved 
and deserves a better understanding. 
In what follows, we assume that in the strong coupling regime of the 
superconformal KW gauge theory, the 
anomalous dimension of twist two operators \eqref{okw} 
indeed scales as $\sqrt{\la}$,
as predicted by the analysis of the sigma model solitons \eqref{elong}.  

In \cite{kn} it was shown that adding $M$ fractional  D3-branes 
(D5-branes wrapping the 2-cycle of the conifold)
breaks the conformal invariance. The resulting 
$\caln=1$ supersymmetric gauge theory has been studied in 
details in\footnote{For a nice review 
see \cite{hko}.} \cite{ks}. Here, the gauge theory on the 
world volume of the branes is $SU(N+M)\times SU(N)$ 
with the same matter content as in the KW gauge theory: 
chiral superfields $A_i$ 
in the $(N+M,\overline{N})$ representation, and chiral superfields 
$B_j$ in $(\overline{N+M},N)$. We also have the same superpotential 
\eqref{sup}. The $M$-deformed theory still 
has $SU(2)\times SU(2)\times U(1)$ global symmetry. 
As argued in \cite{ks}, the sum of the gauge couplings 
\eqref{sum} still remains the exactly marginal direction, 
while the difference of the couplings runs
\begin{equation}
\frac{4\pi}{g_2^2}-\frac{4\pi}{g_1^2}\sim M\ln(\mu/\Lambda)
[3+2(1-\gamma)]\,,
\eqlabel{difrun}
\end{equation}  
where $\gamma$ is the anomalous dimension of operators 
${\rm tr} A_i B_j$, and $\Lambda$ is the strong coupling scale.
As a result of \eqref{difrun}, the $M$-deformed KW theory 
undergoes a series of self-similarity transformations
(a cascade \cite{ks} of Seiberg dualities \cite{s94})
which can be succinctly characterized as if the rank $N$
(the number of regular D3-branes at the tip of the conifold) 
develops an ``anomalous dimension'', so that \cite{bbh}
\begin{equation}
N\to N_{eff}(\mu)\sim g_s M^2 \ln (\mu/\Lambda)\,.
\eqlabel{neff}
\end{equation}
The support for the interpretation \eqref{neff} comes from studies 
of the high temperature thermodynamics of this gauge theory 
\cite{bbh,k2}, and the computation of the correlation 
functions \cite{kr}. The above description of the deformed 
KW gauge theory is clearly physically inadequate in the IR, where
the effective rank \eqref{neff} becomes  negative. 
In \cite{ks}, Klebanov and Strassler analyzed in details the case when 
\begin{equation}
N=1,\qquad  {\rm mod}\ M\,.
\eqlabel{nm}
\end{equation}
They showed that in this case the 
cascade actually stops before reaching $N_{eff}<0$. The physical reason
is rather simple. It turns out, for \eqref{nm}, the gauge theory 
develops a mass gap in the IR, so that in the effective low energy 
description the gauge couplings stop running (because of the 
mass gap there are no charged zero modes), and as a result 
\begin{equation}
N_{eff}(\mu)\equiv1,\qquad \mu<\Lambda\,.
\end{equation}    
With this physical mechanism of stopping the duality 
cascade in mind, it is easy to understand now how to terminate  
Klebanov-Strassler duality cascade in the IR at the
conformal fixed point. Indeed, consider conformal compactification 
of the KW gauge theory on the $S^3$. Such a theory has a mass
gap, and as a result its $M$-deformation (changing the rank of one
of the  gauge groups $N\to N+M$) would stop cascading 
at energy scale $\mu_0$ set by the size of the compactification 
$S^3$. Obviously, it is possible to arrange ``initial''
value $N_{eff}(\mu_0)=N_0\gg 1$, so that the duality cascade would stop 
with $SU(N_0+M)\times SU(N_0)$ gauge theory, which 
can be made arbitrarily close of being conformally 
invariant if $\frac{M}{N_0}\ll 1$. Notice that according to 
\eqref{neff} (see \cite{hko} for a more 
precise statement for $N_{eff}(\mu)$), there is a window of energy scales
\begin{equation}
\mu_0\ll \mu_i<\mu<\mu_f\,,
\eqlabel{window}
\end{equation}
 such that 
\begin{equation}
M\ll g_s M^2 \ln (\mu/\Lambda)\ll N_0\,.
\eqlabel{nwindow}
\end{equation} 
Additionally, we can take 
\begin{equation}
\ln \frac{\mu_f}{\Lambda}\gg 1\,,
\eqlabel{addconst}
\end{equation}
provided $\frac{M^2}{N_0}$ is small enough.
Physically, we need conditions \eqref{window}---\eqref{addconst}
so that the cascading gauge theory is (a) perturbative 
in the appropriate description along the Seiberg duality 
cascade,  (b) being probed at scales much shorter 
than the $S^3$ compactification scale, so that we 
can use the flat-space renormalization group flow 
equations, and (c), though 
at these  energy scales the theory underwent many steps 
of the Seiberg duality cascade, it is still a small 
deformation of the $SU(N_0)\times SU(N_0)$ $\caln=1$ 
superconformal gauge theory.    
The dual supergravity background to the $M$-deformed 
KW gauge theory in the regime  \eqref{window}--\eqref{addconst} 
has been constructed analytically  to the leading 
order in the deformation in \cite{bt}. 

We now turn to the leading correction to the anomalous dimensions
of operators \eqref{okw} due to the $M$-deformation, 
with constraints   \eqref{window}--\eqref{addconst}. 
The perturbative computation \eqref{okwa}  would 
go through with the only change $N\to N_{eff}(\mu)$. 
The dual supergravity computation predicts  
that at large 't Hooft coupling $\la\gg 1$, in the conformally 
invariant KW theory the anomalous dimensions of these operators scale as 
\begin{equation}
\triangle_{\calo_n^{ij}}-(n+2)\sim \sqrt{\la} \ln n\,.
\eqlabel{largeo}
\end{equation}     
Since in the regime \eqref{window}--\eqref{addconst} the deformed 
theory is  almost conformally invariant, we expect that 
the leading correction to \eqref{largeo} would be due 
to\footnote{From the supergravity analysis \cite{bt}, in the KT gauge theory 
on the $S^3$ the string coupling is not constant, but rather 
$g_s\equiv g_s(\mu)$. The scale dependence is actually very mild and 
comes with a factor of $M^2/N_0$. } 
\begin{equation}
\la\to \la_{eff}(\mu)\equiv g_s N_{eff}(\mu)\sim
g_s N_0 \left(1+\frac{M^2}{N_0}\ \ln\frac{\mu}{\Lambda}+M^2\ 
o(\ln(\mu/\Lambda))+ o(M^2)\right)\,,
\eqlabel{ldef}
\end{equation}
thus we expect the leading correction to be  
\begin{equation}
\triangle_{\calo_n^{ij}}-(n+2)\sim \sqrt{\la_0}\ln n+
\frac{\sqrt{\la_0} M^2}{N_0}\ln n \ln\frac{\mu}{\Lambda}\,,
\eqlabel{largeo2}
\end{equation}     
where $\la_0\equiv g_s N_0$.
In the next section we compare  \eqref{largeo2}
with the energy of the highly excited 
string states on the leading Regge trajectory 
in the dual supergravity background \cite{bt}.
We argue that these string states should be thought of
as being dual to gauge theory operators at energy scales 
$\ln\mu/\Lambda \sim \ln n$. In this case the subleading correction 
to the anomalous dimensions of operators \eqref{okw} is predicted 
from \eqref{largeo2} to scale as $M^2/N_0 \ln^2 n$. We rather 
find the leading in $M^2/N_0$ correction to the energy of these 
string states to scale as $M^2/N_0\ln^3 S$, where $S$ is a 
spin of the state.
   
We would like to conclude this section with some 
conjectures about the properties of 
operators  \eqref{okw} in the ultraviolet following from the 
discussion above. 
Clearly, in the UV, whether we compactify KT gauge theory
or not, should be irrelevant. Since in the UV ($\mu\to \infty$), 
$M/N_{eff}(\mu)\to 0$, it is reasonable to assume that 
KT gauge theory approaches
conformal fixed point, which has the properties 
of the standard $\caln=4$ superconformal gauge theory with 
scale dependent number of colors, determined by \eqref{neff}. 
This statement is definitely not new, and is implicit in 
many studies of the KT gauge theory.    
In this case, motivated by \eqref{elong}, we would expect
anomalous dimension of operators \eqref{okw} to be\footnote{Note that 
in the UV the t' Hooft coupling is always large, that's why we should use 
the sigma model result \eqref{elong}.} 
\begin{equation}
\triangle_{\calo_n^{ij}}-(n+2)\sim \sqrt{\la_{eff}(\mu)}\ln n\sim 
g_s^{1/2} M \sqrt{\ln\frac{\mu}{\Lambda}}\ln n\,. 
\eqlabel{vlargeo2}
\end{equation}
Obviously, it makes sense to talk about anomalous dimension 
only when   
\begin{equation}
\sqrt{\la_{eff}(\mu)}\ln n\ll n\, .
\eqlabel{sense}
\end{equation}
Unfortunately, as we explain in  details in the following section, 
the lack of the full nonlinear solution for the dual 
supergravity does not allow us to test \eqref{vlargeo2}.

\section{The supergravity story}
In this section we would like to compare 
\eqref{largeo2} with the dual sigma model computation in
the supergravity background \cite{bt}. 
After reviewing  the construction of the dual 
supergravity to the deformed KW gauge theory, 
we extract the leading Regge trajectory of closed 
strings. In the regime dual to  
\eqref{window}-\eqref{addconst} we 
find subleading correction to the energy  
of highly excited string states to differ from 
that implied by  \eqref{largeo2}. 
We also comment on the difficulty 
studying  ``very long strings'',
which probe the anomalous dimension 
of twist two operators \eqref{okw} far in the 
ultraviolet, expected to be given by 
\eqref{vlargeo2}.
A related work appeared in \cite{t}.

\subsection{SUGRA dual to the KT gauge theory on $S^3$ }

We begin with reviewing the supergravity solution of
\cite{bt} realizing supergravity dual to the KT gauge theory 
compactified on $S^3$. 

The deformed 10-d Einstein frame metric takes the form
\begin{equation}
\begin{split}
ds^2_{E}=&-f_1^{-1/2} dX_0^2+\r f_2^{-1/2}
(dS^3)^2+\frac{d\r^2}{4\r(1-\r)^2}\\ &+f_3^{1/2}
e_{\psi}^2+f_4^{1/2}\left(e_{\theta_1}^2+e_{\phi_1}^2+
e_{\theta_2}^2+e_{\phi_2}^2\right)\,,
\end{split}
\eqlabel{metric}
\end{equation}
with 
\begin{equation}
(dS^3)^2=d\b_1^2+\cos^2\b_1\left(d\b_2^2+\cos^2\b_2 d\f^2\right)\,,
\eqlabel{s3}
\end{equation}
and $e_{\psi}$, $e_{\theta_i}$, $e_{\phi_i}$ are 
the standard $T^{1,1}$ vielbeins (see for example (2.4) of \cite{bt}).
The $p$-form fields are as in the extremal KT solution
\begin{equation}
\begin{split}
F_3=&P e_\psi\wedge \left(e_{\theta_1}\wedge
e_{\phi_1}-e_{\theta_2}\wedge e_{\phi_2}\right),\qquad B_2=P k(\r)
\left(e_{\theta_1}\wedge e_{\phi_1} -e_{\theta_2}\wedge
e_{\phi_2}\right)\,,\\ F_5=&\calf+\star\calf,\qquad \calf=K(\r)
e_\psi\wedge e_{\theta_1}\wedge e_{\phi_1}\wedge e_{\theta_2}\wedge
e_{\phi_2}, \qquad K(\r)=4+2 P^2 k(\r)\,,
\end{split}
\eqlabel{fluxes}
\end{equation}
where the $P=0$ normalization of the five form flux $K$ is such that
the background $AdS_5$ radius is $L=1$. Also we take the bare 
string coupling to be $g_s=1$.  With this normalization, 
$P$ is related to the gauge theory parameters $M,N_0$ of the previous 
section as follows
\begin{equation}
P^2\equiv \frac{M^2}{N_0}\,.
\eqlabel{pmn}
\end{equation}
In \eqref{metric} the radial
coordinate $\r \in [0,1)$ and the warp factors $f_i$ differ by
$O(P^2)$ terms from the $AdS_5\times T^{1,1}$ geometry\footnote{The
standard $AdS_5$ metric in global coordinates $ds_{AdS}=-\cosh^2 r\
dX_0^2+\sinh^2 r\ (dS^3)^2+dr^2 $ is obtained with the identification
$\r=\tanh^2 r$. }
\begin{equation}
\begin{split}
f_1(\r)=&(1-\r)^2+P^2\f_1(\r),\qquad f_2(\r)=(1-\r)^2+P^2 \f_2(\r)\,,\\
f_3(\r)=&1+P^2\f_3(\r),\qquad f_4(\r)=1+P^2 \f_4(\r),\qquad
\Phi(\r)=P^2 \f(\r)\,,
\end{split}
\eqlabel{warpf}
\end{equation}
where $\Phi$ is the dilaton\footnote{The complete nonlinear 
system of differential equations for the warp factors 
$f_i(\r)$, the dilaton $\Phi(\r)\equiv \ln g_s(\r)$, and the 3-form 
flux $k(\r)$ obtained from the type IIB supergravity equations 
of motion is given in Appendix.}.

The analytical solution for the warp factors $\f_i$ and $k(\r)$ to
leading order in $P^2$ was found in \cite{bt}. Here we reproduce only
the IR/UV  ($\r\to 0/\r\to 1_-$) asymptotics.
\begin{equation}
\begin{split}
k(\r)&=\frac{\r}{2}+\frac{\r^2}{4}+O(\r^3),\qquad
\phi(\r)=\frac{\r}{8}+\frac{\r^2}{48}+O(\r^3)\,,\\
\phi_1(\r)&=\left(15\dd-\frac{57}{8}+\frac 23\pi^2\right)\r
+\left(\frac{13}{48}-\frac 53\dd-\frac{2}{27} \pi^2\right)\r^2
+O(\r^3)\,,\\
\phi_2(\r)&=\left(\frac{35}{3}\dd-\frac{67}{12}+\frac{14}{27}\pi^2\right)\r
+\left(-\frac{107}{120}+\dd+\frac{2}{45} \pi^2\right)\r^2 +O(\r^3)\,,\\
\phi_3(\r)&=3\dd+\left(-\frac{31}{8}+12\dd+\frac 13 \pi^2\right)\r +O(\r^2)
\,,\\
\phi_4(\r)&=\left(3\dd-\frac 74+\frac 16\pi^2\right)
+\left(-\frac{25}{4}+12\dd+\frac{7}{12} \pi^2\right)\r +O(\r^2)\,,
\end{split}
\eqlabel{ir}
\end{equation}
as $\r\to 0_+$ and
\begin{equation}
\begin{split}
k(x)&=-\frac{1}{2}\ln x,\qquad
\phi(x)=\left(-\frac 14 +\frac{1}{24}\pi^2 \right)-\frac{1}{4}x 
+O(x^2\ln x)\,,\\
\phi_1(x)&=\left(\frac{1}{16}\ln^2 x-\frac{5}{16}\ln x\right) 
x^2+O(x^2)\,,\\
\phi_2(x)&=\left(\frac{1}{16}\ln^2 x-\frac{5}{16}\ln x\right) 
x^2+O(x^2)\,,\\
\phi_3(x)&=\left(-\frac 14 \ln x+\frac 18\right)+\frac{1}{12}
x+O(x^2)\,,\\
\phi_4(x)&=\left(-\frac 14 \ln x+\frac 18\right)-\frac{1}{24}
x+O(x^2)\,,
\end{split}
\eqlabel{uv}
\end{equation}
as $x\equiv (1-\r)\to 0_+$.
In \eqref{ir} $\delta$ (referred to as $\alpha_1$ in (5.21) of \cite{bt})
has been computed numerically\footnote{From (5.23) and the footnote
(22) of \cite{bt}, $\delta\approx 0.0646108$.}. 
Clearly we can trust ``ultraviolet'' asymptotics \eqref{uv} 
as long as 
\begin{equation}
P^2 \ln^2 x\ll 1\,,
\eqlabel{gravconst}
\end{equation}
which is a gravity dual 
to the gauge theory requirement of staying close to the 
IR conformal fixed point (the second inequality in \eqref{nwindow}).  
Notice that  the supergravity constraint \eqref{gravconst} appears to 
be stronger than the corresponding gauge theory statement, 
once we identify\footnote{This identifications 
comes from comparing the anomalous dimension of the rank \eqref{neff} 
with the dual supergravity statement of the leading $P^2$ radial dependence of 
the 3-form flux: $k(x)=-\frac 12 \ln x$.} 
\begin{equation}
\ln \frac{\mu}{\Lambda}\sim -\ln x\,.
\eqlabel{energyscale}
\end{equation}
A related point (mentioned in \cite{bt}), is that  
the UV asymptotics for $\f_1$, $\f_2$  differ from the corresponding 
asymptotics of the extremal KT solution \cite{kt0002}. 
It is this difference that will be responsible for the 
disagreement of the leading Regge trajectory of the long 
strings with the gauge theory result  \eqref{largeo2}.

Before we move on to study string solitons in the deformed 
KT background, we would like to explain 
one subtlety\footnote{I would like to thank Leo Pando Zayas for raising 
this issue and a very useful discussion.} 
associated with the leading $P^2$ solution of \cite{bt},
giving rise to asymptotics \eqref{ir}, \eqref{uv}. 
This (regular) supergravity solution does not have a 
free parameter apart from the bare string coupling 
(which we set equal to one in the infrared), 
once $P$ is set, and we choose $L=1$. 
Physically, we expect a one parameter family of 
nonsingular KT deformations, represented by 
\begin{equation}
\xi\equiv \frac{\Lambda}{\mu_0}\,, 
\eqlabel{xidef}
\end{equation}
where $\Lambda$ is the strong coupling scale of 
KT gauge theory, and 
$\mu_0$ is the gauge theory compactification scale set by the size of the
$S^3$. We expect that this parameter appears as nonlinear 
effect in $P^2$ in the ``infrared''  of the supergravity 
solution \cite{bt}. Presumably this is due to the fact that, 
in the infrared, the deformation is about the conformal background, 
which is insensitive to a specific value of scale $\mu_0$, as long 
as it is nonzero. While it is possible to identify a candidate 
parameter in the ``ultraviolet'' (where the supergravity 
background is a small deformation of the extremal KT geometry 
\cite{kt0002} ), to conclusively settle this issue one needs 
the full nonlinear solution for the deformed background.
Consider the most general regular in the 
IR (small $\r$) solution to \eqref{fullsystem}. We find\footnote{As in 
\cite{bt}, we set $f_1(\r=0)=1$. This initial value is related to the freedom 
of rescaling the time coordinate, and is also present in KT background 
\cite{kt0002}.}
\begin{equation}
\begin{split}
k(\r)&=\dd_1\ \r+O(\r^2)\,,
\cr
e^{\Phi(\r)}&=1+\frac 14 P^2 \dd_1\ \r+O(\r^2)\,,\cr
f_1&=1-\frac 14 \dd_1 \left(P^2+32\dd_1 \dd_2^{1/2}\right)\ \r +O(\r^2)
\,,\cr
f_2&=1+\frac {1}{18}\left(24 \sqrt{2}\dd_1^{1/2} \dd_2^{1/4}
-128 \dd_1^2 \dd_2^{1/2}-5 \dd_1 P^2 -8 \dd_1 \dd_2 -24\right)\ \r+O(\r^2)
\,,\cr
f_3&=\dd_2-\frac 14 \dd_1\dd_2 \left(32 \dd_1 \dd_2^{1/2}
-16 \dd_2 +3 P^2\right)\ \r+O(\r^2)
\,,\cr
f_4&=\frac 12 \dd_1^{-1} \dd_2^{-1/2}-\frac 18 \dd_1^{-1/2}\dd_2^{-1/2}
\left(8 \dd_1^{1/2}\dd_2 -12 \sqrt{2}\dd_2^{1/4}+32\dd_1^{3/2}\dd_2^{1/2}+P^2 
\dd_1^{1/2}
\right)\ \r+O(\r^2)\,,
\end{split}
\eqlabel{nonlinear}
\end{equation}  
where $\dd_1$, $\dd_2$  are free parameters.
Matching \eqref{nonlinear} with \eqref{ir}, necessary to 
reproduce the leading KT asymptotics in the UV, gives 
\begin{equation}
\begin{split}
\dd_1&=\frac 12+\left(\frac 78 -\frac 94\dd-\frac{1}{12}\pi^2\right)
 P^2+O(P^4)\,,\cr
\dd_2&=1+3\delta P^2+O(P^4)\,.
\end{split}
\eqlabel{match}
\end{equation}
Indeed, both parameters are fixed at order $P^2$. 
As we show now, small deformation of the extremal KT solution in the 
UV  has a single free parameter. To study UV asymptotics ($\r\to 1_-$)
of \eqref{fullsystem}, it is convenient to use the 3-form flux 
function $k(\r)$ as a new radial coordinate $r\equiv k(\r)$. The 
UV asymptotics are then simply $r\to \infty$. Rewriting 
\eqref{fullsystem} for the new radial coordinate,
we find the following asymptotics as $r\to \infty$
\begin{equation}
\begin{split}
g_s&=g_0\left( 1+e^{-2r/g_0} \calg(r)+O(e^{-4r/g_0})\right)\,,\cr
f_1&=\ga \left(1+\frac 12 P^2 r +\frac 18 P^2 g_0\right)
e^{-4r/g_0}\left( 1+e^{-2r/g_0} \calf_1(r)+O(e^{-4r/g_0})\right)\,,\cr
\r^{-2} f_2&=\ga \left(1+\frac 12 P^2 r +\frac 18 P^2 g_0\right)
e^{-4r/g_0}\left( 1+e^{-2r/g_0} r \calf_2(r)+O(e^{-4r/g_0})\right)\,,\cr
f_3&=\left(1+\frac 12 P^2 r +\frac 18 P^2 g_0\right) 
\left( 1+e^{-2r/g_0} \calf_3(r)+O(e^{-4r/g_0})\right)\,,\cr
f_4&=\left(1+\frac 12 P^2 r +\frac 18 P^2 g_0\right) 
\left( 1+e^{-2r/g_0} \calf_4(r)+O(e^{-4r/g_0})\right)\,,
\end{split}
\eqlabel{larger}
\end{equation}
and (compare with \eqref{metric})
\begin{equation}
\frac{(d\r)^2}{4\r(1-\r)^2}\equiv G_{rr} (dr)^2\,,
\eqlabel{radial}
\end{equation}
with 
\begin{equation}
\begin{split}
\calg(r)&=-\frac 14 P^2 g_0 \ga^{1/2}\,,\cr
\calf_1(r)&=\frac{1}{36} P^2 g_0 \ga^{1/2}\
\frac{120 P^2 r +23 P^2 g_0+240}{4 P^2 r+P^2 g_0+8}\,,\cr
\calf_2(r)&=P^2 \ga^{1/2}+\frac{1}{12}\ga^{1/2}\left(13 P^2 g_0+24
\right)\ r^{-1}+O(r^{-2})\,,\cr
\calf_3(r)&=\calf_1(r)-\frac 34 P^2 g_0\ga^{1/2}
\,,\cr
\calf_4(r)&=\calf_1(r)-\frac 78 P^2 g_0\ga^{1/2}\,,
\end{split}
\eqlabel{larder2}
\end{equation}
where $\ga$ is the parameter we conjecture is related to 
$\xi$ in \eqref{xidef}, and $g_0$ is the asymptotic value 
of the string coupling. 
In \eqref{larder2} all functions except $\calf_2$ have been determined
analytically. The function $\calf_2$ is given by 
\begin{equation}
\begin{split}
\calf_2(r)&=\frac{8\ga^{1/2}e^{2 r/g_0}}{9 g_0 r}\biggl(
\alpha+\int^{1/r}\ dx\biggl\{ \frac{e^{-2/(g_0 x)}(P^2+2 x)}{x^3 \left( x (P^2 g_0 
 +8) +4 P^2\right)^2}\cr
&\times \left(
x^2 (11 P^4 g_0^2+78 P^2 g_0+144)+3 P^2 x (13 P^2 g_0+48)+36 P^4
\right)\biggr\}\biggr)\,,
\end{split}
\eqlabel{intrep}
\end{equation}  
with the integration constant $\alpha$ determined from the 
regularity requirement as $r\to \infty$, \eqref{larder2}. 
We could not evaluate \eqref{intrep} in elementary functions. 
Finally,
\begin{equation}
G_{rr}=g_0^{-4} \left(1+\frac 12 P^2 r +\frac 18 P^2 g_0\right)
\left(1+e^{-2r/g_0} g_{rr}(r)+O(e^{-4r/g_0})\right)\,,
\eqlabel{grr}
\end{equation}
with
\begin{equation}
g_{rr}=\frac{\ga^{1/2}}{36}\ \frac{144 P^4 r^2
+12 P^2 (48+19 P^2 g_0) r+41 P^4 g_0^2+
456 P^2 g_0+576}{4  P^2 r +P^2 g_0+8}\,.
\eqlabel{grrs}
\end{equation}
Curiously, even though $g_s$ 
approaches constant both in the IR and in the UV, these constants 
are not the same: from \eqref{ir}, \eqref{uv} we find
\begin{equation}
\ln g_0=\left(-\frac 14 +\frac{1}{24}\pi^2\right) P^2+ O(P^4)\,.
\eqlabel{assg0}
\end{equation}

\subsection{Rotating string in deformed KT geometry}
We now discuss classical string solutions in the background geometry
\eqref{metric}.

We take sigma model action  in the conformal gauge\footnote{Note 
that $G_{ij}$ is the string frame metric.}
\begin{equation}
\begin{split}
S=\frac{1}{4\pi\a'}\int d^2\sigma G_{ij} \del_\a X^i \del^\a X^j\,.
\end{split}
\eqlabel{action}
\end{equation}
This action is supplemented by the constraints
\begin{equation}
\begin{split}
G_{ij}& \left(\del_\tau X^i \del_\tau X^j+\del_\sigma X^i \del_\sigma X^j\right)=0\,,\\
G_{ij}& \del_\tau X^i \del_\sigma X^j=0\,.
\end{split}
\eqlabel{const}
\end{equation}
Consider a closed string rotating in the $(\r,\phi)$ plane in the 
deformed  KT background \eqref{metric}
\begin{equation}
\begin{split}
\r&=\r(\s),\qquad \f=\w \tau,\qquad X_0=k\t\,,\\
\b_i&=0,\qquad \theta_i=0,\qquad  \f_i=0,\qquad \psi=0\,,
\end{split}
\eqlabel{adsrot}
\end{equation}
for  constants $(k,\w)$. Equations of motion and the constraints are satisfied provided
\begin{equation}
\begin{split}
0&=\del_\s\left(G_{\r\r} \del_\s \r \right)
-\frac 12 \del_\r\left(G_{\r\r}\right) (\del_\s \r)^2+
\frac 12\del_\r \left(G_{\f\f} \w^2- G_{X_0 X_0} k^2\right)\,,\\
0&=G_{\r\r} (\del_\s \r)^2+\left(G_{\f\f} \w^2- G_{X_0 X_0} k^2\right)\,,
\end{split}
\eqlabel{ec}
\end{equation}
where 
\begin{equation}
\begin{split}
G_{X_0 X_0}&=e^{\Phi/2}\ f_1^{-1/2}\,,\\
G_{\f\f}&=e^{\Phi/2}\ \r f_2^{-1/2}\,,\\
G_{\r\r}&=e^{\Phi/2}\ \frac{1}{4\r (1-\r)^2}\,. \\
\end{split}
\eqlabel{gdef}
\end{equation}
The space-time energy is given by
\begin{equation}
E=\frac{k}{2\pi\a'}\int_0^{2\pi} d\s\ G_{X_0X_0}\,,
\eqlabel{energy}
\end{equation}
and the spin is 
\begin{equation}
S=\frac{\w}{2\pi\a'}\int_0^{2\pi} d\s\ G_{\f\f}\,.
\eqlabel{spin}
\end{equation}
We consider the simplest ``one-fold'' string configuration where 
the interval $0\le \s <2\pi$ is split into 4 segments: for  $0< \s <\pi/2$
the function $\r(\s)$ increases from $0$ to its maximum value $\r_0$
such that\footnote{Primes  denote derivatives with respect to
$\s$.} $(\r'(\pi/2)=0)$
\begin{equation}
0=G_{\f\f}(\r_0) \w^2- G_{X_0 X_0}(\r_0) k^2\,,
\eqlabel{rho0}
\end{equation}
then for $\pi/2<\s<\pi$ decreases to zero, etc. The periodicity 
of $\s$ implies additional condition on the parameters
\begin{equation}
2\pi=\int_0^{2\pi}d\s=4\int_0^{\r_0} d\r \sqrt{\frac{G_{\r\r}}{k^2 G_{X_0X_0}-
\w^2 G_{\f\f}}}\,.
\end{equation}
In what follows we introduce 
\begin{equation}
\frac{\w^2}{k^2}=1+\eta\,.
\eqlabel{eta}
\end{equation} 

\subsubsection{Short strings}
A short string limit corresponds to 
\begin{equation}
\eta\gg 1\,.
\eqlabel{short}
\end{equation}
From \eqref{rho0} we find 
\begin{equation}
\r_0\approx \frac 1\eta\ll 1\,.
\eqlabel{sizesmall}
\end{equation} 
In this case the rotating string is hardly stretched, 
so we can replace the complicated deformed KT geometry 
\eqref{metric} with almost flat space. 
We expect  the leading Regge trajectory to be 
a small deformation of that  in flat space.
Indeed, a  somewhat tedious but straightforward computation gives
\begin{equation}
\begin{split}
2\pi\a' E&=\frac{1}{\e^{1/2}}\ _2F_1 \left(-\frac 12,
\frac 12 ,1,-\frac {1}{\e}\right)\\
&-\frac {|P|}{\e} \left(\frac{20}{3}\dd-\frac{37}{12}+\frac{8}{27}\pi^2
\right)^{1/2}\left(1+\frac{1}{\eta}\left(\frac{44\dd-831/40+
88\pi^2/45}{40\dd-37/2+16\pi^2/9}\right)+O\left(\frac{1}{\e^2}\right)\right)\\
&-\frac{\pi P^2}{\e^{3/2}}\left(\frac{5}{18}\pi^2-3+\frac{25}{4}\dd\right)
\left(1+\frac{1}{\e}\left(
\frac{22\pi^2/9-267/10+55\dd}{16\pi^2/9-96/5+40\dd}\right)
+O\left(\frac{1}{\e^2}\right)\right)\\
&+O\left(P^3\right)\,,
\end{split}
\eqlabel{energys}
\end{equation}
for the  energy, and 
\begin{equation}
\begin{split}
2\pi\a' S&=\frac{\sqrt{1+\e}}{2\e^{3/2}}\ _2F_1 \left(\frac 12,
\frac 32 ,2,-\frac {1}{\e}\right)\\
&-\frac {|P|}{\e^{3/2}} 
\left(\frac{20}{3}\dd-\frac{37}{12}+\frac{8}{27}\pi^2
\right)^{1/2}
\left(1+\frac{1}{\eta}\left(
\frac{24\dd-461/40+16\pi^2/15}{40\dd-37/2+16\pi^2/9}\right)
+O\left(\frac{1}{\e^2}\right)\right)\\
&-\frac{\pi P^2}{\e^{2}}\left(\frac{5}{24}\pi^2-\frac 94+\frac{75}{16}
\dd\right)
\left(1+\frac{1}{\e}\left(
\frac{86\pi^2/45-1047/50+43\dd}{8\pi^2/5-432/25+36\dd}\right)
+O\left(\frac{1}{\e^2}\right)\right)\\
&+O\left(P^3\right)\,,
\end{split}
\eqlabel{spins}
\end{equation}
for the spin.
Furthermore, we find
\begin{equation}
\frac{E^2}{2\a' S}=\left(1+O\left(\frac{1}{\e}\right)\right)
+\frac{|P|}{\e^{3/2}}\left(\a_1+O\left(\frac{1}{\e}\right)\right) 
+\frac{P^2}{\e^2}\left(\a_2+O\left(\frac{1}{\e}\right)\right)\,,
\eqlabel{ratios}
\end{equation}
where $\a_i$ are some constants easily computable from 
\eqref{energys},\eqref{spins}.
As in \cite{gkp0204}, reintroducing the scale $L$, 
\eqref{ratios}
reproduces the standard operator-state correspondence of the
AdS/CFT duality.

\subsubsection{Long strings}
The long strings correspond to 
\begin{equation}
\e\ll 1\,.
\eqlabel{long}
\end{equation}
Here a priori we have to  consider two cases
\begin{equation}
P^2 (\ln\e)^2\ll 1\,,
\eqlabel{case1}
\end{equation}
which corresponds to the small deformation of the 
conformal infrared fixed point, or 
\begin{equation}
P^2 (\ln\e)^2\gg 1\,,
\eqlabel{case2}
\end{equation}
for the ``very long strings'' corresponding to 
the UV of the KT gauge theory.
It is the first regime, \eqref{case1} that is of most interest.
Using the asymptotics \eqref{uv}, from \eqref{rho0} we find 
\begin{equation}
\r_0\approx\frac{1}{1+\eta}+O(P^4)\,.
\eqlabel{rlarge}
\end{equation}
From \eqref{energy}, \eqref{spin} we find 
\begin{equation}
\begin{split}
2\pi\a' \left(E-S\right)&=2 
\int_0^{\r_0}\ d\r\ \frac{(1-(1+\eta)^{1/2}\r)}{\r^{1/2}(1-\r)^{3/2}
(1-(1+\eta)\r)^{1/2}} \left(1-\frac{P^2}{64}\ln^2(1-\r)+\cdots\right)
\end{split}
\eqlabel{longregge}
\end{equation}
where $\cdots$ denotes subleading in $P^2$ correction to the 
one indicated.  Notice that 
in the limit $\eta\to 0$ (we still have to satisfy \eqref{case1}!)
the integral diverges. The divergence comes solely from the 
upper limit, and thus justifies the use of \eqref{uv}. 
We interpret this ``localization'' of the 
divergence as the statement that the
stretched rotating string ``probes'' anomalous 
dimension of the dual gauge theory operators
at energy scale dual to its radial extent. Using gauge/gravity 
renormalization group relation \eqref{energyscale} this scale is  
\begin{equation}
\ln \frac{\mu}{\Lambda}\sim -\ln(1-\r_0)\sim -\ln \eta\,.
\eqlabel{mueta}
\end{equation} 
Carefully extracting the divergent as $\eta\to 0$ 
part of \eqref{longregge} we find
\begin{equation}
2\pi (\epsilon-s)\equiv 2\pi\a' (E-S)=\left(-2\ln \eta+o\left(\ln\eta\right)\right)
+P^2 \left(\frac{1}{96} (\ln\eta)^3+o\left((\ln\eta)^3\right)\right)\,.
\eqlabel{leading}
\end{equation}
We also need to relate $\eta$ and $s$. From \eqref{spin},
\begin{equation}
 2\pi s\equiv 2\pi\a' S= \left(\frac {4}{\e}+\ln\e+O(\e\ln\e)\right)-\frac{P^2}{16 \e}
\left((\ln\e)^2+O(\ln\e)\right)+O(P^3)\,,
\end{equation} 
so that 
\begin{equation}
\e= \frac{2}{\pi s}\left(1-\frac{\ln s}{2\pi s}+\cdots\right)
\left(1-P^2\left\{\frac{ (\ln s)^2}{64}+\cdots\right\}\right),
\eqlabel{eta1}
\end{equation}
where again we kept only the leading terms.
With \eqref{eta1}, we arrive at our final expression 
for the energy of closed string states on the 
leading Regge trajectory in the deformed KT geometry
\begin{equation}
\ep-s=\left(\frac{1}{\pi}\ln\frac{s\pi}{2}+O\left(\frac{\ln s}{s}\right)\right)
+\frac{P^2}{2\pi}\left(\left(\ln\frac{s\pi}{2}\right)^3+
O\left((\ln s)^2\right)\right)\,.
\eqlabel{finenergy}
\end{equation} 
Two comments are in order. First,   
\eqref{finenergy}  does not reproduce the leading correction to the 
anomalous dimension of operators \eqref{okw} 
at energy scales specified in \eqref{window}-\eqref{addconst}:
with the standard gauge/gravity dictionary \eqref{dic}
the agreement would imply 
\begin{equation}
\ep-s=\left(\frac{1}{\pi}\ln\frac{s\pi}{2}+O\left(\frac{\ln s}{s}\right)\right)
+\frac{P^2}{2\pi}\left(A \left(\ln s\right)^2+
O\left((\ln s)^2\right)\right)\,,
\eqlabel{finenergyex}
\end{equation} 
for some constant $A$. 
This is rather puzzling, as the twist two operators \eqref{okw}
are the natural candidate dual, motivated by the translation of the 
GKP analysis to the deformed KW conformal gauge theory. 
Second, we would have found the agreement for the leading 
$P^2$ scaling, had the UV asymptotics of the deformed 
KT geometry \eqref{uv} contained single logarithms,
rather than $\ln^2 x$. As claimed in \cite{bt}, it is 
the single logarithm asymptotics  for $\phi_1$ and $\phi_2$ 
of \eqref{uv}, that is required for the precise 
agreement with the UV of the extremal KT geometry. In \eqref{uv} 
we extended  the asymptotic analysis of \cite{bt} 
and confirmed the $x^2 \ln^2 x$ leading asymptotic. 
We did not manage to find a  different from \cite{bt}
regular deformation of the KW background by the 
3-form fluxes of the KT type.

\subsubsection{Very long strings}
In the previous section, using only asymptotic linear in $P^2$  
geometry of the deformed KT background we determined the leading 
Regge trajectory of the long strings \eqref{case1}.
The question we would like to address here is whether 
using the far ultraviolet asymptotics of the deformed 
KT geometry \eqref{larger}--\eqref{grrs},
we can still identify very long rotating strings, \eqref{case2},
with operators \eqref{okw} in the far ultraviolet. 
The latter are expected to have anomalous dimensions
given by \eqref{vlargeo2}. 
 
Unfortunately, we can not do so in a computationally controllable way.
The main difficulty stems from the fact that 
while the energy/spin of the rotating string in case 
\eqref{case1} was  dominated by the contribution from 
the ``stretched end'' (so that the use of \eqref{uv} for the 
background geometry was indeed justified), in case 
\eqref{case2} we find energy/spin to be very sensitive to the 
infrared region of the geometry. Furthermore, 
while the infrared geometry is completely 
regular for the asymptotics \eqref{uv}, it is unphysical 
(singular) for \eqref{larger}.

To give some more details let's compare the expression 
for the energy of the ``long'' \eqref{case1}, and 
the ``very long'' \eqref{case2} string. 
To illustrate the point, in the case of long strings it is 
sufficient to set $P=0$. The gravity background 
is then simply $AdS_5$, and  the energy \eqref{energy} of a rotating
string (we use $\rho$ as a radial coordinate ) is        
\begin{equation}
2\pi\a' E_{long}=2 \int_0^{\r_0=1/(1+\eta)}\ \frac{d\r}{\r^{1/2}(1-\r)^{3/2}
(1-(1+\eta)\r)^{1/2}}\,. 
\eqlabel{fel}
\end{equation}   
Indeed, as $\eta\to 0$ the integral \eqref{fel} is dominated at 
the upper bound, and diverges as $\eta^{-1}$.
In other words, in this case the main contribution does come from the 
ends of the folded stretched string. As the latter stretches almost to the
boundary of the global $AdS_5$, we are justified to 
use asymptotics \eqref{uv} for the $P^2$ deformation. 
Note also that these leading UV asymptotics are regular 
for small $\r$ (we have to set $x\to 1_-$ in \eqref{uv}),
thus the contribution from the lower bound in \eqref{fel}
is still small.

Consider now the energy of the very long string. 
Here, using the asymptotics  
\eqref{larger}  and keeping the leading terms in the 
integrand we find
\begin{equation}
2\pi\a' E_{very\ long}\approx
4\sqrt{2} g_0^{1/2}\ga^{-1/4}\ \int_{ r_{ir}/g_0}^{r_0/g_0}\ dt\ 
\frac{e^t}{\left(g_0 P^2 \ga^{1/2}\ t e^{-2 t}-2\eta\right)^{1/2}}\,,
\eqlabel{fevl}
\end{equation}
where we had to introduce the infrared cutoff $r_{ir}$ 
to make sense of \eqref{fevl}, also 
\begin{equation}
\frac 12 r_0 e^{-2 r_0/g_0} P^2 \ga^{1/2}\equiv \eta\,.
\eqlabel{roeta}
\end{equation} 
Notice that unlike \eqref{fel}, the integral \eqref{fevl}
 crucially depends on the infrared cutoff $r_{ir}$.
It is possible that  the problems with the long strings are 
due to a bad choice of radial coordinate. Recall that here 
we used the 3-form flux dependence $k(\r)$ as a radial coordinate 
of the deformed KT geometry. It is also possible that 
these very long strings are unsuitable probes for the 
operators \eqref{okw} in the far UV.  
This is an interesting open question to pursue further.

\section{Summary}
In this paper we discussed  the gauge/string correspondence 
when gauge theory is formulated in curved space. 
We argued that in an appropriate regime, the cascading 
gauge theory on the world-volume of regular 
and fractional D3-branes is a small deformation 
about the infrared conformal fixed point.
Following the GKP prescription 
we attempted to reproduce the anomalous dimension 
of certain twist two operators from the dual nonlinear 
sigma model computation in the supergravity background 
of \cite{bt}. We did not succeed in doing this. 

We will now summarize the main steps of the gauge/string correspondence 
studied in this paper. 
\nxt 
We assumed the validity of the GKP identification 
of $\caln=4$ $SU(N)$ supersymmetric Yang-Mills theory 
operators \eqref{o} with the long strings spinning 
in $AdS_5$. The leading Regge trajectory of these highly 
excited string states then provides a  prediction
for the anomalous dimensions of twist two operators \eqref{o}
at large 't Hooft coupling.  
\nxt
It follows then that the GKP proposal would go 
through for the Klebanov-Witten $\caln=1$ 
$SU(N)\times SU(N)$ superconformal
gauge theory \cite{kw}, with twist two operators 
now being \eqref{okw}. This is intuitively clear 
once we recall that the KW gauge theory is an infrared 
fixed point of the  $\zet_2$ 
orbifold of the $\caln=4$ gauge theory, deformed by the 
mass term for the adjoint chiral superfields in the 
$\caln=2$ vector multiplets.  
By (literally) repeating GKP analysis, we 
 then have a  prediction for the anomalous dimension 
of operators \eqref{okw} at large 't Hooft coupling. 
\nxt 
Next, we considered deformation of the KW gauge theory, 
when one of the ranks of the two gauge groups is shifted 
$N\to N+M$. We assumed that the resulting gauge theory would 
still be almost conformally invariant if $M/N\ll 1$. 
As shown in \cite{ks}, while classically we 
can think of $N$ and $M$ as being some constants, this is 
inconsistent quantum mechanically: the deformation $N\to N+M$
breaks conformal invariance, and under the renormalization group 
flow this theory very fast becomes strongly coupled.
The proper (perturbative) quantum-mechanical description of this 
deformed theory is in terms of ``cascading'' gauge theory
$SU(N(\mu))\times SU(N(\mu)+M)$ 
where the rank $N$ becomes scale dependent.
In the ultraviolet, the cascade goes forever as 
$N(\mu)\to \infty$ as $\mu/\Lambda\to \infty$, 
while in the infrared it stops dynamically 
with $N(\mu/\Lambda\to 0)< M $. Thus in 
the original KT model, the cascading gauge theory is 
far from being conformally invariant in the infrared.
While, it is tempting to say that KT gauge theory is almost ``conformal'' 
in the ultraviolet because $M/N(\mu)\to 0$, 
the precise meaning 
of this statement is not clear. 
\nxt
A way to terminate the duality cascade in the IR with 
a conformal gauge theory was proposed in \cite{bt}. 
The idea is simply to conformally compactify KW 
gauge theory, and  then deform it by the shift 
$N\to N+M$. While there is still the Seiberg duality cascade 
(induced by the RG flow) in this compactified gauge theory, 
which still continues forever in the UV, the termination 
of the cascade in the IR is rather different: because of the 
kinematic mass gap in the theory (due to the compactification)
the RG flow stops below the scale of the lightest charged 
states, in turn stopping the duality cascade. We called 
this energy scale $\mu_0$.  
\nxt 
We then explained the constraints on the parameters of the 
model, namely $N_0\equiv N(\mu_0)$, $M$, and the range of energy scales 
such that the deformed cascading gauge theory is 
close to the superconformal $\caln=1$ $SU(N_0)\times SU(N_0)$
gauge theory. First, since the cascading 
KT gauge theory  always has gauge group ranks 
differing by the multiple of 
$M$, we should choose $N_0\gg M$. Second, in order to use the 
flat-space cascading picture of Klebanov and Strassler, 
\cite{ks}, we should be able to study the theory 
at energy scales $\mu$ much higher than the compactification scale
$\mu\gg \mu_0$. One the other hand, the energy scales 
of interest should not be too high, if we still want this cascading
theory to be close to the infrared superconformal 
theory:  $N(\mu)-N_0\ll N_0$. From the gauge theory 
perspective all these conditions 
are mutually compatible for small enough $M^2/N_0$, and the 
appropriate range of energy scales. 
\nxt
Accepting above steps, it is natural to assume 
(and we did this in a paper) that the dominant
correction to the anomalous dimension of operators 
\eqref{okw} for large 't Hooft coupling $\lambda=g_s N_0\gg 1$
due to the deformation $N_0\to N_0+M$ for one of the gauge groups
rank, would come from simply replacing $\sqrt{\lambda_0}$ 
in the (dual gravitational) 
expression for the leading Regge trajectory of closed 
string in $AdS_5$ with $\sqrt{\lambda_{eff}(\mu)}$,
properly accounting for the Seiberg duality cascade in the 
theory. 
\nxt 
We proposed to test the latter prediction by studying 
closed stings spinning in the supergravity 
background \cite{bt}. This background was constructed 
as a leading (in $P^2\equiv M^2/N_0$) 
regular deformation of the global 
$AdS_5\times T^{1,1}$ geometry (dual to the 
superconformal $SU(N_0)\times SU(N_0)$) gauge theory 
by turning on certain 3-form fluxes on $T^{1,1}$, 
dual to the gauge theory shift deformation $N_0\to N_0 +M$.
\nxt 
As in the $AdS_5$ space-time, we found that 
the closed string states on the leading Regge 
trajectory with very high spin, were realized by strings 
stretched almost to the boundary of the ($P^2$ deformed)
global $AdS_5$. Dominant contribution 
for the energy/spin of such states was coming from the 
region close to the boundary. We proposed to identify these 
stretched strings as gravity dual to operators \eqref{okw},
probed at energy scales dual to the radial extent of 
these stretched stings.
This energy scale was determined unambiguously by 
identifying the radial dependence of the 3-form flux in the 
dual supergravity with the gauge theory equation
for the scale dependence of the effective rank of the 
cascading gauge theory $N(\mu)$. 
\nxt
We found  different order $P^2$ correction to the
Regge trajectory of long strings in the background \cite{bt}, compare with   
the $M^2/N_0$ corrected anomalous dimension of KW operators 
\eqref{okw}. We traced the difference to the fact that  
the asymptotics of the (linear in $P^2$) 
supergravity solution of \cite{bt}  does not
reproduce precisely the asymptotics of the 
extremal Klebanov-Tseytlin background \cite{kt0002}.

We did not find a physically satisfactory 
explanation for the discrepancy between the long strings
Regge trajectory in the background \cite{bt} 
and the (supposedly gauge theory dual) anomalous dimension of operators 
 \eqref{okw}. One possibility is that there is a different  
from \cite{bt} deformation of the KW supergravity 
background (\ie\ $AdS_5\times T^{1,1}$), 
also regular in the limit of vanishing 
3-form flux. We searched for such solution, still within the 
ansatz of only radial deformation as in \cite{bt}, 
and did not find one. 
Another possible explanation is related to the 
subtlety  of the background \cite{bt}, discussed in
section 3. The point is that the leading $P^2$ deformation 
of \cite{bt} does not contain parameter $\mu_0$. It is thus 
not clear whether the gauge theory requirement $\mu\gg \mu_0$
is actually satisfied  before 
$P^2 \ln^2 x $ in  Eq.~\eqref{uv} becomes much larger then one, 
invalidating the use of these asymptotics for the study 
of long spinning strings relevant to operators \eqref{okw}
with anomalous dimensions given by  \eqref{largeo2}.  
To resolve the latter issue it is necessary either to construct 
the next order correction in $P^2$ to the background 
\cite{bt} (to see the appearance of $\mu_0$, and verify that 
$P^2 \ln^2 \mu_0\ll 1$  is indeed possible to arrange),
or to understand the renormalization group flow of the 
compactified cascading gauge theory at energy scales $\mu\sim \mu_0$.
We hope to report on this  in the future.

\section*{Acknowledgments}
It is a pleasure to thank Igor Klebanov, Asad Naqvi,
Leo Pando Zayas and Arkady Tseytlin for interesting and stimulating
discussions.  
This work has been  supported in part by the U.S.~Department of 
Energy.

\section*{Appendix}
Type IIB supergravity equations of motion for the background 
\eqref{metric}, \eqref{fluxes}
reduce to the following system
\begin{equation}
\begin{split}
0&=\left(\frac{(k')^4 f_3\ \r^8 (1-\r)^4}{f_1 f_2^3\ g_s^4 }\right)'
-\frac{2(k')^3 f_3^{1/2}\ (2+ P^2 k)\ \r^7 (1-\r)^2}{f_1 f_2^3 f_4\ g_s^3}
\,,\cr
\cr
0&=\left(\frac{(g_s')^4 f_3 f_4^4\ \r^8 (1-\r)^4}{
f_1 f_2^3\ g_s^4}\right)'+\frac{P^2 (g_s')^3 f_4^3\ \r^7(1-\r)^2}{f_1 f_2^3
\ g_s^4}\left(4 (k')^2 f_3\ \r (1-\r)^2-f_3^{1/2} g_s^2 \right)\,,\cr
\cr
0&=\left(\frac{(f_4')^4 f_3\ \r^8 (1-\r)^4}{f_1 f_2^3}\right)'
+\frac{(f_4')^3\ \r^7(1-\r)^2}{f_1 f_2^3 f_4\ g_s} \biggl(
4 g_s f_3^{1/2}\left(2+P^2 k\right)^2+P^2 g_s^2 f_3^{1/2} f_4\cr
&+4 P^2 \left(k'\right)^2 f_3 f_4\ \r (1-\r)^2
-8 g_s f_3 f_4\left(3 f_4^{1/2}-f_3^{1/2}\right)
\biggr)\,,\cr
\cr
0&=\left(\frac{(f_3')^4 f_4^4\ \r^8(1-\r)^4}{f_1 f_2^3 f_3^3}\right)'
+\frac{(f_3')^3 f_4^2\ \r^7 (1-\r)^2}{f_1 f_2^3 f_3^{5/2}\ g_s}\biggl(
4 g_s \left(2+P^2 k\right)^2\cr
&+g_s f_4 \left(3 P^2 g_s-16 f_3\right)
-4 P^2 \left(k'\right)^2 f_3^{1/2} f_4\ \r (1-\r)^2  
\biggr)\,,\cr\cr
0&=\left(\frac{(f_1')^4 f_3 f_4^4\ \r^8(1-\r)^4}{f_1^5 f_2^3}\right)'
+\frac{(f_1')^3 f_4^2\ \r^7(1-\r)^2}{f_1^4 f_2^3\ g_s}\biggl(
4 g_s f_3^{1/2} (2+ P^2 k)^2\cr
&+P^2 g_s^2 f_3^{1/2} f_4
+4 P^2 \left(k'\right)^2 f_3 f_4\ \r (1-\r)^2 
\biggr)\,,\cr
\cr
0&=\left(\frac{(f_2')^4 f_3 f_4^4\ \r^{14}(1-\r)^4}{f_1 f_2^7}\right)'
-\frac{2 (f_2')^3\left(\frac{f_3 f_4^4}{f_1}\right)'\ \r^{13}(1-\r)^4}{f_2^6}
\cr
&+\frac{(f_2')^3 f_4^2\ \r^{12}(1-\r)^2}{f_1 f_2^{13/2}\ g_s}\biggl(
4 P^2 \left(k'\right)^2 f_2^{1/2} f_3 f_4\ \r^2 (1-\r)^2+
4 g_s f_2^{1/2} f_3^{1/2} \left(2+P^2 k\right)^2 \ \r\cr 
&+g_s f_4\biggl(
P^2 g_s f_2^{1/2}f_3^{1/2}\ \r
+8 f_2^{1/2}f_3 f_4 \left(2 r-1\right)\left(1-\r
\right) +8 f_2 f_3 f_4 \biggr)
\biggr)\,,
\end{split}
\eqlabel{fullsystem}
\end{equation}
where  primes denote derivatives with respect to $\r$. 
There is also a first order constraint (consistent with 
\eqref{fullsystem}) coming from gauge fixing 
the radial coordinate in \eqref{metric}.

\end{document}